\documentclass[12pt,fleqn]{article}

\usepackage{amsmath,amssymb}
\usepackage{graphicx}
\usepackage{xcolor}
\usepackage{physics}
\usepackage[top=0.75in, bottom=0.75in, left=0.725in, right=0.725in]{geometry}
\usepackage[small]{caption}
\usepackage[noblocks]{authblk}
\usepackage{hyperref}
\usepackage{cite}
\usepackage{comment} 
\usepackage{xcolor}
\usepackage{ulem}

\newcommand{\eg}{\textit{e.g.,~}}
\newcommand{\ie}{\textit{i.e.,~}}

\newcommand{\gev}{\ensuremath{\,\mathrm{GeV}}}

\begin{document}

\title{\sf QCD bounds on leading-order hadronic vacuum polarization contributions to the muon anomalous magnetic moment
}

\author[1]{Siyuan Li\thanks{siyuan.li@usask.ca}}
\author[1]{T.G. Steele\thanks{tom.steele@usask.ca}}
\author[2]{J. Ho\thanks{jason.ho@dordt.edu}}
\author[3]{R. Raza\thanks{rraza@tru.ca}}
\author[3]{K. Williams\thanks{williamsk16@mytru.ca}}
\author[3]{R.T. Kleiv\thanks{rkleiv@tru.ca}}

\affil[1]{Department of Physics and Engineering Physics, University of Saskatchewan, Saskatoon, SK, S7N~5E2, Canada}\affil[2]{Department of Physics, Dordt University, Sioux Center, Iowa, 51250, USA}
\affil[3]{Department of Physics, Thompson Rivers University, Kamloops, BC, V2C~0C8, Canada}

\maketitle

\begin{abstract}
QCD bounds on the leading-order (LO) hadronic vacuum polarization (HVP) contribution to the anomalous magnetic moment of the muon  ($a_\mu^{\mathrm{HVP,LO}}$, $a_\mu=\left(g-2\right)_\mu/2$) are determined by imposing H\"older inequalities and related inequality constraints on systems of Finite-Energy QCD sum-rules. This novel methodology is complementary to lattice QCD and data-driven approaches to determining $a_\mu^{\mathrm{HVP,LO}}$. For the light-quark ($u,d,s$) contributions up to five-loop order in perturbation theory in the chiral limit, LO in light-quark mass corrections, next-to-leading order in dimension-four QCD condensates,  and to LO in dimension-six QCD condensates, we find that $\left(657.0\pm 34.8\right)\times 10^{-10}\leq a_\mu^{\mathrm{HVP,LO}} \leq \left(788.4\pm 41.8\right)\times10^{-10}\,$, bridging the range between lattice QCD and data-driven values.
\end{abstract}

\newpage

\section{Introduction} \label{sec:intro}
In the summer of 2023, the Muon $g-2$ experiment  at Fermilab announced an updated result to the measurement of $a_\mu \equiv (g-2)_\mu/2$, increasing the precision of their previous measurement by a factor of two \cite{PhysRevLett.131.161802} (see also \eg Ref.~\cite{Muong-2:2024hpx}). This updated experimental result reinforces the tension between experimental measurements and predictions from the Standard Model using data-driven and dispersive methods, pushing the disagreement between this new experimental observation and the prediction from theory \cite{AOYAMA20201} up to $5.0\sigma$ \cite{PhysRevLett.131.161802}. In addition to this new experimental evidence, recent precision measurements of the pion form factor by CMD-3 have been used to calculate the lowest-order hadronic contributions to $a_\mu$ 
\cite{CMD-3:2023rfe}, and found agreement with \cite{PhysRevLett.131.161802} to within $0.9\sigma$. Furthermore, a recent calculation by the Budapest–Marseille–Wuppertal (BMW) collaboration using lattice QCD (LQCD) reached sub-percent levels of precision competitive with data-driven and dispersive methods \cite{Borsanyi:2020mff}. This high-precision LQCD calculation of $a_\mu$ is in significantly better agreement with current experimental measurements. While efforts are ongoing by the LQCD community to produce new calculations of sub-percent precision \cite{Kuberski:2023qgx}, the results of the BMW collaboration produced a new tension between theoretical methods.

Currently, contributions to $a_\mu$ from the hadronic vacuum polarization (HVP) dominate the uncertainties in the Standard Model calculation. In the data-driven approach, the leading-order (LO) dispersion integral for the contributions to $a_\mu$ from HVP (\ie $a_\mu^{\mathrm{HVP,LO}}$) is given by \cite{AOYAMA20201,PhysRev.174.1835,Steele1991}
\begin{equation}
    a_\mu^{\mathrm{HVP, LO}} = \frac{1}{4\pi^3} \int_{4m_\pi^2}^{\infty}\sigma^{H}(t) K(t)\,\mathrm{d}t  
    \label{eq:a_HVP}
\end{equation}
where $\sigma^{H}$ is the $e^+e^-$ to hadrons cross section and $K(t)$, the kernel function, is given by
\begin{equation}
    K(t) = \int_{0}^{1} dx\, \frac{x^2(1-x)}{x^2+(1-x)t/m_\mu^2}\,,
    \label{eq:K_exact}
\end{equation}
where $m_\mu$ is the muon mass.
Using the hadronic $R$-ratio 
\begin{gather}
R(t)=\frac{\sigma^{H}\left(t\right)}{\sigma\left(e^+e^-\rightarrow\,\mu^+\mu^-\right)}\,,
\label{eq:Rratio}
\end{gather}
with
\begin{gather}
\sigma\left(e^+e^-\rightarrow\,\mu^+\mu^-\right)=\frac{4\pi\alpha^2}{3t^2}\left(t+2m_\mu^2\right)\sqrt{1-\frac{4m_\mu^2}{t}}=\frac{4\pi\alpha^2}{3t}+{\cal O}\left(\frac{1}{t^3}\right)\,,
\label{eq:lepton_sigma}
\end{gather}
where $\alpha$ is the fine-structure constant,  Eq.~\eqref{eq:a_HVP} can be expressed as
\begin{equation}
 a_\mu^{\mathrm{HVP,LO}}= \frac{\alpha^2}{3\pi^2}\int_{4m_\pi^2}^{\infty}\frac{1}{t}R(t)K(t)\,\mathrm{d}t\,,
 \label{eq:a_HVP_2}
\end{equation}
where the approximation associated with \eqref{eq:lepton_sigma} is negligible. Since the hadronic $R$-ratio can be expressed in terms of the hadronic vacuum polarization spectral function $R(t) = 12\pi \mathrm{Im}\Pi^H(t)$ \cite{Caprini:1984ud,Steele1991}, a QCD expression for Eq.~\eqref{eq:a_HVP} can be written in terms of the hadronic spectral function $\mathrm{Im}\Pi^H(t)$,
\begin{equation}
    a_\mu^{\mathrm{QCD}} = \frac{4\alpha^2}{\pi}\int_{4m_\pi^2}^{\infty}\frac{1}{t}\mathrm{Im}\Pi^H(t)K(t)\,\mathrm{d}t\,.
    \label{eq:a_QCD}
\end{equation}
We can relate \eqref{eq:a_HVP_2}  and \eqref{eq:a_QCD} to QCD sum-rule methods by approximating Eq.~\eqref{eq:K_exact} as 
\begin{equation}
    K(t)\approx \frac{m_\mu^2}{3t}=K_{\rm approx}(t)
\label{eq:K_approx}
\end{equation}
to obtain
\begin{equation}
    a_\mu^\mathrm{QCD} \approx \frac{4 m_\mu^2\alpha^2}{3\pi} \int_{4m_\pi^2}^\infty \frac{1}{t^2} \mathrm{Im}\Pi^H\left(t\right) \mathrm{d}t\,,
    \label{eq:a_mu_QCD}
\end{equation}
where the effects of the approximation associated with \eqref{eq:K_approx} will be discussed below. The challenges of a QCD determination of $a_\mu^{\mathrm{HVP,LO}}$ arise from the $1/t^2$ behaviour in \eqref{eq:a_mu_QCD} that emphasizes the low-energy region.  

QCD sum-rules \cite{Shifman:1978bx,Shifman:1978by} (see \eg  \cite{Reinders:1984sr,Narison:2002woh,Gubler:2018ctz,Colangelo:2000dp} for reviews) implement quark-hadron duality by relating a QCD prediction to an integrated hadronic spectral function, and hence \eqref{eq:a_mu_QCD} suggests the possibility of using QCD sum-rules for determining $a_\mu^{\mathrm{HVP,LO}}$. In particular, the structure of \eqref{eq:a_mu_QCD} is such that it can be written in terms of a finite-energy QCD sum rule (FESR) defined by \cite{Floratos:1978jb,Hubschmid:1980rm,Bertlmann:1984ih,Caprini:1985ex}
\begin{equation}
F_{k}\left(s_{0}\right) = \int_{t_0}^{s_0} \frac{1}{\pi}\mathrm{Im}\Pi^H \left(t\right) t^k\, \mathrm{d}t\,,
\label{eq:fesr}
\end{equation}
where $k$ is an integer that indicates the weight of the sum-rule and $t_0$ is a physical threshold. In \eqref{eq:fesr}, the left-hand side is obtained from a QCD prediction, and hence the FESRs relate a QCD prediction to an integrated hadronic spectral function. Writing \eqref{eq:a_mu_QCD} in terms of \eqref{eq:fesr} gives
\begin{equation}
    a_\mu^\mathrm{QCD} \approx \frac{4 m_\mu^2\alpha^2}{3} F_{-2}\left(\infty\right) \ge \frac{4 m_\mu^2\alpha^2}{3} F_{-2}\left( s_0\right) \,.
    \label{eq:a_mu_QCD_tom}
\end{equation}
In the last step of Eq.~\eqref{eq:a_mu_QCD_tom},  positivity of the hadronic spectral function has been used to obtain a lower bound.  As outlined below, the presence of the parameter $s_0$ allows optimization of our theoretical prediction. 
Unfortunately, determining a field-theoretical expression for $F_{-2}\left(s_0\right)$ requires knowledge of low-energy constants, and hence a direct theoretical prediction is not possible. Various QCD sum-rule approaches have been used to circumvent this issue (see \eg Refs.~\cite{Caprini:1984ud,Casas:1985yw,Steele1991,Narison:2023srj}). In this paper we examine the fundamental properties of the field theoretical result \eqref{eq:a_mu_QCD_tom} through the application of the H\"older, Cauchy-Schwarz, and related inequalities  to obtain QCD lower and upper bounds on the LO hadronic vacuum polarization contribution to the anomalous magnetic moment of the muon $a_\mu^{\mathrm{HVP,LO}}$. 

In Section~\ref{sec:fesr} the fundamental inequalities for lower and upper bounds are developed. Section~\ref{fesr_section_tom} provides the necessary  QCD expressions and input parameters  
for light-quark ($u,d,s$) contributions up to five-loop order in perturbation theory in the chiral limit, LO in light-quark mass corrections, next-to-leading order (NLO) in dimension-four QCD condensates,  and to LO in dimension-six QCD condensates.  Analysis methodology and results for $a_\mu^{\mathrm{HVP,LO}}$ are presented in Section~\ref{analysis_sec_tom}, and the Appendix updates
the Laplace sum-rule bounds on  $a_\mu^{\mathrm{HVP,LO}}$ in Ref.~\cite{Steele1991} with current determinations of the necessary QCD input parameters, five-loop perturbative corrections, and NLO dimension-four condensate contributions.

\section{QCD Finite-Energy Sum Rule Bounds on $a_\mu^{\mathrm{QCD}}$}\label{sec:fesr}

\subsection{Lower Bounds}
H\"older inequalities have previously been developed for QCD Laplace \cite{Benmerrouche:1995qa} and Gaussian sum-rules \cite{Ho:2018cat}, and their application can be used to constrain the region of sum-rule parameter space in the study of hadronic systems (see \eg~Refs.~\cite{Benmerrouche:1995qa,Ho:2018cat,Steele:1998ry,Shi:1999hm,Wang:2016sdt,Yuan:2017foa}).  Extending this H\"older inequality methodology to FESRs allows us to establish fundamental bounds on the theoretically-undetermined FESR $F_{-2}\left(s_0\right)$, leading to a constraint on $a_\mu^\mathrm{QCD}$ via \eqref{eq:a_mu_QCD_tom}.
 
 The H\"older inequality  
 is expressed generally as \cite{Beckenbach:1961, Berberian:1965}
\begin{equation}\label{eq:holder_general}
\abs{\int_{t_1}^{t_2}f\left(t\right)g\left(t\right)\mathrm{d}\mu} \leq 
\left( \int_{t_1}^{t_2}\abs{f\left(t\right)}^{p}\mathrm{d}\mu \right)^{\frac{1}{p}}
\left( \int_{t_1}^{t_2}\abs{g\left(t\right)}^{q}\mathrm{d}\mu \right)^{\frac{1}{q}},\quad \frac{1}{p} + \frac{1}{q} = 1 \,.
\end{equation}
With careful choice  of functions $f(t)$, $g(t)$, and using positivity of $\mathrm{Im}\Pi^H\left(t\right)$ to define the measure $\mathrm{d}\mu=\frac{1}{\pi}\mathrm{Im}\Pi^H\left(t\right)\mathrm{d}t$ our H\"older inequality becomes
\begin{align}
\abs{\int_{t_0}^{s_0}t^{\alpha+\beta} \frac{1}{\pi}\mathrm{Im}\Pi^H\left(t\right) \mathrm{d}t} \leq &
\left( \int_{t_0}^{s_0}\abs{t^{\alpha}}^{p}\frac{1}{\pi}\mathrm{Im}\Pi^H \left(t\right) \mathrm{d}t \right)^{\frac{1}{p}} \left( \int_{t_0}^{s_0}\abs{t^{\beta} }^{q}\frac{1}{\pi}\mathrm{Im}\Pi^H\left(t\right) \mathrm{d}t \right)^{\frac{1}{q}}.
\label{eq:fesr_tom}
\end{align}
Because the QCD quantity $F_k\left(s_0\right)$ in Eq.~\eqref{eq:fesr} must inherit the properties associated with the hadronic spectral function, Eq.~\eqref{eq:fesr_tom} can be expressed in terms of the FESRs 
\begin{equation}
    F_{\alpha + \beta}\left(s_{0}\right) \leq \left[F_{\alpha\, p}\left(s_{0}\right)\right]^{\frac{1}{p}} \left[F_{\beta \,q}\left(s_{0}\right)\right]^{\frac{1}{q}} \;
    \rightarrow  \; F_{\alpha + \beta} \left(s_{0}\right)\leq \left[F_{\alpha \,p}\left(s_{0}\right)\right]^{\frac{1}{p}} \left[F_{ \frac{\beta\,p}{p-1}}\left(s_{0}\right)\right]^{\frac{p-1}{p}} \,.
    \label{eq:fesr-holder}
\end{equation}
Eq.~\eqref{eq:fesr-holder} results in a family of inequalities which can be used to place a lower bound on $F_{-2}\left(s_0\right)$. These are restricted due to the conditions from the H\"older inequality 
[Eq.~\eqref{eq:holder_general}], as well due to the requirement from FESRs $F_k\left(s_0\right)$ that the weight $k$ be an integer. By restricting our attention to inequalities that give a lower bound on $F_{-2}\left(s_0\right)$ through a combination of positive-weight FESR expressions, we derive   the following inequalities:
 \begin{gather}
    F_{-2} \geq\frac{F_{0}^2}{F_{2}}\,, \label{eq:constraint1}\\
    F_{-2} \geq\frac{F_{0}^3}{F_{1}^{2}}\,,  \label{eq:constraint2}\\
    F_{-2} \geq \frac{F_{1}^4}{F_{2}^{3}} \label{eq:constraint3}\,,
\end{gather}
where we have suppressed the $s_0$ dependence in each FESR. These inequalities place a lower bound on $F_{-2}\left(s_0\right)$ through a combination of FESRs that have weights low enough ($0\leq k \leq 2$) to avoid dependence on unknown higher dimension QCD condensates as outlined below.   

Having determined the lower bounds 
\eqref{eq:constraint1}--\eqref{eq:constraint3}, we next determine which is the strongest restriction on $F_{-2}$. Starting from Eq.~\eqref{eq:fesr-holder}, 
we apply the substitutions $\alpha = \frac{k+1}{2}$ and $\beta = \frac{k-1}{2}$, and consider the Cauchy-Schwarz inequality (\ie  the H\"older inequality in Eq.~\eqref{eq:fesr-holder} with $p = q = 2$). This gives
\begin{gather}
    F_k \leq F_{k+1}^{1/2} F_{k-1}^{1/2} \; \rightarrow \; F_k^2 \leq F_{k+1} F_{k-1}\,.
    \label{eq:Cauchy-Schwarz}
\end{gather}
Rearranging this gives us a relationship between ratios of FESRs,
\begin{equation}\label{eq.C-S_for_s0}
    \frac{F_k}{F_{k+1}} \leq \frac{F_{k-1}}{F_k}\,.
\end{equation}
Applying this to our constraints \eqref{eq:constraint1}-\eqref{eq:constraint3}, we find the following hierarchy,
\begin{equation}
F_{-2} \geq \frac{F_{0}^3}{F_1^2} \geq \frac{F_{0}^2}{F_2} \geq \frac{F_1^4}{F_{2}^3} \,.
    \label{eq:fesr-hierarchy}
\end{equation}
The most restrictive lower bound on $F_{-2}\left(s_0\right)$ is therefore provided by 
\begin{equation}
    F_{-2} \geq \frac{F_{0}^3}{F_1^2} \,.
\label{eq:fesr_best_tom}
\end{equation}
From this, taking Eqs.~\eqref{eq:a_mu_QCD_tom} and \eqref{eq:fesr_best_tom},
we can relate this inequality to a bound on $a_\mu^\mathrm{QCD}$,
\begin{equation}
   a_\mu^\mathrm{QCD}
   \geq \frac{4 m_\mu^2\alpha^2}{3} \frac{F_{0}^3\left(s_0\right)}{F_1^2\left(s_0\right)} \,.
   \label{eq:a_mu_QCD-fesr}
\end{equation}

\begin{figure}[htb]
\centering
    \includegraphics{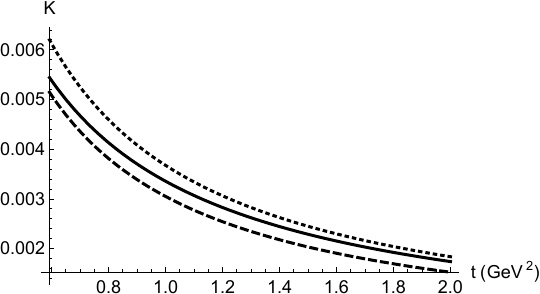}
    \caption{The exact $K(t)$ (solid line)  compared to the approximate form $K_\xi(t)$ with $\xi=0.83$ (lower dashed line) and with $\xi=1$ (upper dotted line).}
    \label{K_approx_fig}
\end{figure}

In obtaining the lower bound  \eqref{eq:a_mu_QCD-fesr} on $a_\mu^\mathrm{QCD}$, the approximation in Eq.~\eqref{eq:K_approx} has been used.
The resulting lower bound  \eqref{eq:a_mu_QCD-fesr} is only valid if this approximation is also a lower bound on $K(t)$. However, the approximation \eqref{eq:K_approx} provides an upper bound on $K(t)$, and $K_{\mathrm{approx}}(t)$ must therefore be re-scaled by a factor $\xi$ to obtain a valid lower bound
\begin{equation}
 K_\xi(t)=\xi K_{\mathrm{approx}}(t)=\xi  \frac{m_\mu^2}{3t}\,.
 \label{K_xi}
\end{equation}
The crucial energy region for determining $\xi$ is the low-energy region from threshold to the $\rho,\omega$  peak. A naive Breit-Wigner $\sigma^{\mathrm{BW}}$ for the $\rho,\omega$  is non-zero at threshold and    provides an overestimate of $\sigma^H(t)$ in the low-energy region. Thus $\xi$ can be  determined by the constraint 
\begin{equation}
\int_{4m_\pi^2}^{m_\rho^2}K(t) 
\sigma^{\mathrm{BW}}(t)\,\mathrm{d}t
\ge \int_{4m_\pi^2}^{m_\rho^2}K_\xi(t) \sigma^{\mathrm{BW}}(t)\,\mathrm{d}t \,.
\label{xi_constraint}
\end{equation}
The inequality \eqref{xi_constraint} is saturated by $\xi=0.83$, 
and as shown in Fig.~\ref{K_approx_fig}, this value of $\xi$  also results in a lower bound   $K_\xi(t)\le K(t)$  beyond the $\rho,\omega$ peak,
\begin{equation}
\int_{4m_\pi^2}^{\infty}K(t) 
\sigma^{H}(t)\,\mathrm{d}t
\ge \int_{4m_\pi^2}^{\infty}K_\xi(t) \sigma^{H}(t)\,\mathrm{d}t \,.
\label{final_xi_inequality}
\end{equation}
Hence  \eqref{eq:a_mu_QCD-fesr} is modified to our final form
\begin{equation}
 a_\mu^\mathrm{QCD} \ge  \xi\frac{4 m_\mu^2\alpha^2}{3} \frac{F_{0}^3\left(s_0\right)}{F_1^2\left(s_0\right)}   \,,~\xi=0.83\,.
   \label{eq:a_mu_QCD-fesr_xi}
\end{equation}
It should be noted from Fig.~\ref{K_approx_fig} that the approximate form $K_\xi(t)$ clearly underestimates the exact $K(t)$ above the $\rho,\omega$ peak, and hence the final bound in Eq.~\eqref{eq:a_mu_QCD-fesr_xi} is expected to be a conservative lower bound. Finally, the utility of the  parameter $s_0$  appearing in \eqref{eq:a_mu_QCD-fesr_xi} is now evident, because it can be varied to find the strongest possible QCD bound.

\subsection{Upper Bounds}
Because the kernel $K(t)$ decreases monotonically with increasing energy and $K(t)<K_{\rm approx}(t)$ (see Fig.~\ref{K_approx_fig}), the following  upper bound can be obtained from \eqref{eq:a_QCD} and \eqref{eq:a_mu_QCD}
\begin{equation}
     a_\mu^\mathrm{QCD} \le \frac{4 m_\mu^2\alpha^2}{3\pi} \int_{t_0}^\infty \frac{1}{t^2}  \mathrm{Im}\Pi^H\left(t\right) \mathrm{d}t\le  \frac{4 m_\mu^2\alpha^2}{3\pi} \frac{1}{t_0^2} \int_{t_0}^\infty  \mathrm{Im}\Pi^H\left(t\right) \mathrm{d}t\,,~t_0=4m_\pi^2\, .
\end{equation}
However, this bound can be improved by adapting and extending the techniques outlined in Ref.~\cite{Dalfovo:1992dr}. Ultimately, the goal is to construct an upper bound on $F_{-2}\left(s_0\right)$, but we illustrate the method of Ref.~\cite{Dalfovo:1992dr} with the necessary step of an upper bound on $F_{-1}\left(s_0\right)$ via the following relation based on positivity of the hadronic spectral function
\begin{equation}
    \int\limits_{t_0}^{s_0}\frac{1}{t}\left[1+A t\right]^2 \mathrm{Im}\Pi^H\left(t\right) \,\mathrm{d}t\ge 0\,.
\end{equation}
By extremizing $A$ to obtain the most stringent relation we find 
\begin{gather}
    F_{-1}\le F_{-1}^{(B)}=\frac{F_0}{t_0}-\frac{\left(F_1/t_0-F_0\right)^2}{\left(F_2/t_0-F_1\right)}\,,
    \label{fm1_B}
    \\
    F_2/t_0-F_1>0\,,
    \label{fm1_B_condition}
\end{gather}
where the FESR dependence on $s_0$ has been suppressed and the subsidiary condition \eqref{fm1_B_condition} is required for the validity of \eqref{fm1_B}.  

An upper bound on $F_{-2}$ can then be obtained by extremizing the relation 
\begin{equation}
    \int\limits_{t_0}^{s_0}\frac{1}{t^2}\left[1+A t\right]^2 \mathrm{Im}\Pi^H\left(t\right) \,\mathrm{d}t\le \frac{1}{t_0} \int\limits_{t_0}^{s_0}\frac{1}{t}\left[1+A t\right]^2 \mathrm{Im}\Pi^H\left(t\right) \,\mathrm{d}t\,,
\end{equation}
to find
\begin{gather}
    F_{-2} \le \frac{F_{-1}^{(B)}}{t_0}-\frac{\left(F_0/t_0-F_{-1}^{(B)}\right)^2}{\left(F_1/t_0-F_0\right)}\,,
   \label{fm2_B_1} 
    \\
    F_1/t_0-F_0>0\,, \left(F_0/t_0-F_{-1}^{(B)}\right)^2<\left(F_0/t_0-F_0^2/F_1\right)^2\,,
    \label{fm2_B_condition_1}
\end{gather}
where the inequality 
$F_{-1} \ge  F_0^2/F_1$ 
[see \eqref{eq:Cauchy-Schwarz}] has been 
used as part of the 
 subsidiary condition  \eqref{fm2_B_condition_1} for the validity of \eqref{fm2_B_1}.
An alternative upper bound on $F_{-2}$ can be obtained by extremizing 
\begin{equation}
    \int\limits_{t_0}^{s_0}\frac{1}{t^2}\left[1+A t\right]^2 \mathrm{Im}\Pi^H\left(t\right) \,\mathrm{d}t\le \frac{1}{t_0^2} \int\limits_{t_0}^{s_0}\left[1+A t\right]^2 \mathrm{Im}\Pi^H\left(t\right) \,\mathrm{d}t
\end{equation}
to obtain
\begin{gather}
     F_{-2} \le F_0/t_0^2-\frac{\left(F_1/t_0^2-F_{-1}^{(B)}\right)^2}{\left(F_2/t_0^2-F_0\right)}\,,
   \label{fm2_B_2}  
   \\
  F_2/t_0^2-F_0>0\,, \left(F_1/t_0^2-F_{-1}^{(B)}\right)^2<\left(F_1/t_0^2-F_0^2/F_1\right)^2 \,,
  \label{fm2_B_condition_2}
\end{gather}
where 
the inequality 
$F_{-1} \ge  F_0^2/F_1$  
[see \eqref{eq:Cauchy-Schwarz}] has again been 
used as part of the 
 subsidiary condition  \eqref{fm2_B_condition_2} for the validity of \eqref{fm2_B_2}. 

 Thus the upper QCD bound that is complimentary  to the lower bound \eqref{eq:a_mu_QCD-fesr} is

\begin{equation}
   a_\mu^\mathrm{QCD}
   \le \frac{4 m_\mu^2\alpha^2}{3} 
   \begin{cases}
   F_{-1}^{(B)}/{t_0}-\frac{\left(F_0/t_0-F_{-1}^{(B)}\right)^2 }{F_1/t_0-F_0}
   \\[12pt]
    F_0/t_0^2-\frac{\left(F_1/t_0^2-F_{-1}^{(B)}\right)^2}{F_2/t_0^2-F_0}
   \,
   \end{cases}  
   ,
   \label{eq:a_mu_QCD-fesr_upper}
\end{equation}
where either \eqref{fm2_B_1} or  \eqref{fm2_B_2} is used for a QCD upper bound on $F_{-2}$. 
Both forms lead to identical numerical values despite the distinct pathways used to obtain them.
Note that similar to the  lower bound  on $F_{-2}$ in \eqref{eq:fesr_best_tom},  the  $F_{-2}$ upper bounds \eqref{fm2_B_1} and \eqref{fm2_B_2} all depend on the well-determined QCD FESRs $\{F_0,F_1,F_2\}$, and similarly the parameter $s_0$ can be varied  to find the strongest possible QCD bound. Combining \eqref{eq:a_mu_QCD-fesr_xi} and \eqref{eq:a_mu_QCD-fesr_upper}, our $a_\mu^\mathrm{QCD}$ bounds emerging from fundamental QCD sum-rule inequalities are
\begin{equation}
 \xi\frac{4 m_\mu^2\alpha^2}{3} \frac{F_{0}^3\left(s_0\right)}{F_1^2\left(s_0\right)} \le   
 a_\mu^\mathrm{QCD}
   \le \frac{4 m_\mu^2\alpha^2}{3} 
   \begin{cases}
   F_{-1}^{(B)}/{t_0}-\frac{\left(F_0/t_0-F_{-1}^{(B)}\right)^2 }{F_1/t_0-F_0}
   \\[12pt]
    F_0/t_0^2-\frac{\left(F_1/t_0^2-F_{-1}^{(B)}\right)^2}{F_2/t_0^2-F_0}
   \end{cases} \,,~\xi=0.83\,,  
  \label{eq:a_mu_QCD-fesr_summary}  
\end{equation}
where the parameter $s_0$ can be  varied independently on both sides of the inequality to find the strongest possible bounds.

\section{Finite-Energy Sum-Rules: QCD Inputs }
\label{fesr_section_tom}
To generate a bound on $a_\mu^\mathrm{QCD}$ from the FESRs in Eq.~\eqref{eq:a_mu_QCD-fesr_summary}, correlation functions 
for the light quark vector current $j_\mu(x)  = \bar{q}(x)\gamma_\mu q(x)$ provide the QCD prediction related  to the hadronic spectral function in \eqref{eq:fesr}.
The original LO calculation of the QCD correlation function $\Pi(Q^2)$ up to dimension-six in the operator-product expansion \cite{Shifman:1978bx,Shifman:1978by,Reinders:1984gu} 
(see also Refs.~\cite{Reinders:1984sr,Pascual1984,Narison:2002woh,Steele1991})
has been extended to NLO in the dimension-four QCD condensates \cite{Surguladze:1990sp,Wang:2016sdt,Chetyrkin:1985kn} and 
$\overline{\mathrm{MS}}$-scheme perturbative contributions up to five loop order in the chiral limit \cite{PhysRevLett.101.012002,Gorishnii:1990vf,Surguladze:1990tg,Chetyrkin:1996ez,Chetyrkin:1979bj,Dine:1979qh,Celmaster:1979xr} 
(see also Refs.~\cite{PhysRevD.67.034017,dEnterria:2022hzv}) 

\begin{align}
    \Pi\left(Q^2\right) =   & \frac{1}{4\pi^2} \Pi^\mathrm{pert}\left(Q^2\right)
    - \frac{3 m_q^2(\nu)}{2\pi^2 Q^2} + 2\langle m_q \bar{q}q\rangle\frac{1}{Q^4}\left(1+\frac{1}{3}\frac{\alpha_s(\nu)}{\pi}\right)
    \nonumber \\
    & +\frac{1}{12\pi}\langle \alpha_s G^2 \rangle \frac{1}{Q^4}
    \left(1+\frac{7}{6}\frac{\alpha_s(\nu)}{\pi}\right)-\frac{224}{81}\pi \alpha_s \langle \bar{q}\bar{q}qq\rangle \frac{1}{Q^6}\,.
    \label{eq:correlator}
\end{align}
In addition, $\Pi\left(Q^2\right)$ also requires an additional pre-factor of the quark charge $Q_q^2$. The perturbative contributions in \eqref{eq:correlator} are given by
\begin{gather}
  \frac{1}{\pi}\mathrm{Im}\Pi^\mathrm{pert}\left(t,\nu\right) =S\left[x(\nu),L(\nu)\right] = 1+\sum_{n=1}^{\infty} x^{n} \sum_{m=0}^{n-1} T_{n,m}L^m\,,
    \label{eq:pertseries}
    \\
    x(\nu) \equiv \frac{\alpha_s(\nu)}{\pi}\,,~L(\nu) \equiv \log\left(\frac{\nu^2}{t}\right)\,,
\end{gather}
where the coefficients $T_{n,m}$ given in Table \ref{table:coeff_nf_3_and_4} are implicitly a function of  $N_f$, the number of active quark flavours. As outlined below, the energy range in our analysis results in a renormalization scale appropriate to $N_f=3$ and $N_f=4$. The QCD parameters necessary for Eqs.~\eqref{eq:fesr} and \eqref{eq:correlator} are listed in Table \ref{table:parameters}. 

\begin{table}[ht]
\centering
\renewcommand{\arraystretch}{1.5}
\begin{tabular}{|c|c|c|c|c||c|c|c|c|c|}
\hline
 $N_f = 4$ & $m = 0$ & $m = 1$ & $m = 2$ & $m = 3$ 
 & $N_f = 3$ & $m = 0$ & $m = 1$ & $m = 2$ & $m = 3$\\ \hline
 $n = 1$ & 1 & -- & -- & --
 &$n = 1$ & 1 & -- & -- & --\\ \hline
 $n = 2$ & 1.52453 & 25/12 & -- & -- 
 &$n = 2$ & 1.63982 & 9/4 & -- & -- \\ \hline
 $n = 3$ & -11.6856 & 9.56054 & 625/144 & --
 &$n = 3$ & -10.2839 & 11.3792 & 81/16 & -- \\ \hline
 $n = 4$ & -92.91 & -56.90 & 36.56 & $\frac{15625}{1728}$ 
 &$n = 4$ & -106.896 & -46.2379 & 47.4048 & 729/64 \\ \hline
\end{tabular}
\caption{$\overline{\mathrm{MS}}$-scheme coefficients 
$T_{n,m}$ within \eqref{eq:pertseries} for the imaginary part of the vector-current correlation function up to five-loop order for $N_f = 4$ (left) and $N_f = 3$ (right). The four-loop results are given in Ref.~\cite{PhysRevD.67.034017}, the five-loop coefficient $T_{4,0}$ is from \cite{PhysRevLett.101.012002}, and five-loop logarithmic coefficients $T_{4,1}$, $T_{4,2}$, and $T_{4,3}$ are generated from the renormalization group analysis of Ref.~\cite{PhysRevD.67.034017} via the four-loop ($N_f=4$ and $N_f=3$) $\overline{\mathrm{MS}}$-scheme $\beta$ function \cite{vanRitbergen:1997va}.}
\label{table:coeff_nf_3_and_4}
\end{table}

\begin{table}[htb]
\centering
\renewcommand{\arraystretch}{1.3}
\begin{tabular}{c|c|c}
\hline
\multicolumn{1}{|c|}{Parameter} & \multicolumn{1}{c|}{Value} & \multicolumn{1}{c|}{Source} \\ \hline
 $\alpha$ & $1/137.036$ & \cite{PDG2022} \\
 $\alpha_s\left(M_\tau\right)$ & $0.312\pm 0.015$ & \cite{PDG2022} \\
 $m_u(2\,\mathrm{GeV})$ & $2.16^{+0.49}_{-0.26}\,\mathrm{MeV}$ & \cite{PDG2022}\\
 $m_d(2\,\mathrm{GeV})$ & $4.67^{+0.48}_{-0.17}\,\mathrm{MeV}$ & \cite{PDG2022}\\
 $m_s(2\,\mathrm{GeV})$ & $\left(0.0934^{+0.0086}_{-0.0034}\right)\gev$ & \cite{PDG2022} \\
 $f_\pi$ & $\left(0.13056\pm 0.00019\right)/\sqrt{2}$ GeV& \cite{PDG2022} \\
 $m_n \langle \bar{n}n\rangle$ & $-\frac{1}{2}f_\pi^2 m_\pi^2$ & \cite{Gell-Mann:1968hlm}  \\
 $m_s \langle \bar{s}s\rangle$  & $r_m r_c m_n \langle\overline{n}n\rangle $  & \cite{Harnett2021} \\
 $r_c \equiv \langle \bar{s}s\rangle/\langle \bar{n}n\rangle$ & $0.66 \pm 0.10 $  & \cite{Harnett2021} \\
 $m_s/m_n = r_m$ & $27.33^{+0.67}_{-0.77}$ & \cite{PDG2022}  \\
 $\langle \alpha G^2 \rangle$ & $\left(0.0649\pm 0.0035\right)\gev^4$ & \cite{Albuquerque:2023bex}\\
  $\kappa$ & $3.22\pm 0.5$ & \cite{Albuquerque:2023bex} \\
 $ \alpha_s \langle\bar{n}n\rangle^2$ & $\kappa\left(1.8\times 10^{-4}\right)\gev^6$  & \cite{Harnett2021} \\
  $ \alpha_s \langle \bar{s}s\rangle^2$ & $r_c^2 \alpha_s \langle \bar{n}n\rangle^2$ & \cite{Harnett2021}
\end{tabular}
\caption{QCD parameters and uncertainties used in our analysis. Here, $m_n = \left(m_u+m_d\right)/2$ and $\langle\bar{n}n\rangle = \langle\bar{u}u\rangle = \langle\bar{d}d\rangle$.
 }
\label{table:parameters}
\end{table}

The FESR defined via \eqref{eq:fesr} are now constructed up to five-loop order in perturbation theory in the chiral limit, LO in light-quark mass corrections, next-to-leading order (NLO) in dimension-four QCD condensates, and to LO in dimension-six QCD condensates.
Using standard FESR methodology \cite{Floratos:1978jb,Hubschmid:1980rm,Bertlmann:1984ih,Caprini:1985ex}, the resulting
FESR  $F_k$ for weights $k=\{0, 1, 2\}$ as needed for analysis of \eqref{eq:fesr-hierarchy}, \eqref{fm2_B_1}, and \eqref{fm2_B_2}  are given by
\begin{align}
    F_0\left(s_0\right) =  \frac{1}{4\pi^2}\Bigg[1 & + \frac{\alpha_s(\nu)}{\pi}T_{1,0} + \left(\frac{\alpha_s(\nu)}{\pi}\right)^2\left(T_{2,0}+T_{2,1}\right)+ \left(\frac{\alpha_s(\nu)}{\pi}\right)^3\left(T_{3,0}+T_{3,1}+2T_{3,2}\right) \nonumber\\ 
    & + \left(\frac{\alpha_s(\nu)}{\pi}\right)^4\left(T_{4,0}+T_{4,1}+2T_{4,2}+6T_{4,3}\right)\Bigg]s_0 
    -\frac{3}{2\pi^2}  m_q(\nu)^2\,,
    \label{eq:fesr-k-0}
\end{align}
\begin{align}
 F_1\left(s_0\right) =  \frac{1}{8\pi^2}\Bigg[1 & + \frac{\alpha_s(\nu)}{\pi}T_{1,0} + \left(\frac{\alpha_s(\nu)}{\pi}\right)^2\left(T_{2,0}+\frac{1}{2}T_{2,1}\right)+ \left(\frac{\alpha_s(\nu)}{\pi}\right)^3\left(T_{3,0}+\frac{1}{2}T_{3,1}+\frac{1}{2}T_{3,2}\right) \nonumber\\ 
    & + \left(\frac{\alpha_s(\nu)}{\pi}\right)^4\left(T_{4,0}+\frac{1}{2}T_{4,1}+\frac{1}{2}T_{4,2}+\frac{3}{4}T_{4,3}\right)\Bigg]s_0^2 
    \nonumber\\
    &
    - 2\langle m_q \bar{q}q\rangle\left(1+\frac{1}{3}\frac{\alpha_s(\nu)}{\pi}\right)
    - \frac{1}{12\pi}\langle \alpha_s G^2 \rangle\left(1+\frac{7}{6}\frac{\alpha_s(\nu)}{\pi}\right)\,,
    \label{eq:fesr-k-1}
\end{align}
\begin{align}
    F_2\left(s_0\right) =  \frac{1}{12\pi^2}\Bigg[1 & + \frac{\alpha_s(\nu)}{\pi}T_{1,0} + \left(\frac{\alpha_s(\nu)}{\pi}\right)^2\left(T_{2,0}+\frac{1}{3}T_{2,1}\right)+ \left(\frac{\alpha_s(\nu)}{\pi}\right)^3\left(T_{3,0}+\frac{1}{3}T_{3,1}+\frac{2}{9}T_{3,2}\right) \nonumber\\ 
    & + \left(\frac{\alpha_s(\nu)}{\pi}\right)^4\left(T_{4,0}+\frac{1}{3}T_{4,1}+\frac{2}{9}T_{4,2}+\frac{2}{9}T_{4,3}\right)\Bigg]s_0^3 
    - \frac{224}{81} \pi \alpha_s \langle \bar{q}\bar{q}qq\rangle  \,.
    \label{eq:fesr-k-2}
\end{align}
Implicit in Eqs.~\eqref{eq:fesr-k-0}--\eqref{eq:fesr-k-2} is a renormalization scale of $\nu = \sqrt{s_0}$ in both $\alpha_s$ and the running quark masses (see \eg Refs.~\cite{Floratos:1978jb,Hubschmid:1980rm,Bertlmann:1984ih,Caprini:1985ex}).
This can be understood as arising from the renormalization-group equation  satisfied by \eqref{eq:pertseries}
\begin{equation}
    \left(-t\frac{\partial}{\partial t}+\beta(\alpha_s)\frac{\partial}{\partial \alpha_s}\right)\mathrm{Im}\Pi^\mathrm{pert}\left(t,\nu\right)=0\,,
\label{eq:RG_corr}
\end{equation}
where the canonical and anomalous mass dimensions are zero for the vector current.  
From \eqref{eq:RG_corr} it follows that the FESRs satisfy the following renormalization-group equation
\begin{gather}
    \left(-s_0\frac{\partial}{\partial s_0}+\beta(\alpha_s)\frac{\partial}{\partial \alpha_s}+(k+1) \right)F_k^{\rm pert}\left(s_0,\nu\right)
=0 \,,
\label{eq:RG_fesr}
\\
F_k^{\rm pert}\left(s_0,\nu\right) =\int_0^{s_0}   t^k \frac{1}{\pi}\mathrm{Im}\Pi^\mathrm{pert}\left(t,\nu\right)\,dt\,.
\label{pert_fesr}
\end{gather}
Thus apart from the trivial $s_0^{k+1}$ canonical dimension prefactor, the solution of the renormalization-group equation for the QCD expressions \eqref{eq:fesr-k-0}--\eqref{eq:fesr-k-2} is obtained by the standard replacement $\nu^2=s_0$.  For renormalization-group behaviour of the dimension-four NLO contributions, it is helpful to recall that   $\langle m_q \bar{q}q\rangle$ and $\langle \beta G^2\rangle+4 \gamma\langle  m_q \bar{q}q\rangle$ are renormalization-group invariant (see \eg Ref.~\cite{Pascual1984}).

In particular, because we are working to ${\cal O}\left(\alpha_s^4\right)$ in perturbation theory, we numerically solve the  renormalization-group equation 
using the four-loop  $\overline{\mathrm{MS}}$-scheme $\beta$ function \cite{vanRitbergen:1997va}  with $N_f$ appropriate to the active flavours below $s_0$ and using $\alpha_s\left(M_\tau\right)$ as a boundary condition.  For the running quark mass corrections, only the  LO 
($\overline{\mathrm{MS}}$-scheme)
anomalous mass dimension is needed. As outlined below, this $s_0$ energy region will span the range covered by $N_f=4$ and $N_f=3$. We do not implement flavour threshold matching conditions \cite{Chetyrkin:1997un} (see \eg Ref.~\cite{Steele:1998qls} for an example implementation) because such effects are insignificant compared to other sources of theoretical uncertainty. Finally,  the generic light-flavour FESRs \eqref{eq:fesr-k-0}--\eqref{eq:fesr-k-2} require a pre-factor of their quark charge  (\ie~$Q_u^2 = 4/9$ and $Q_d^2 = Q_s^2 = 1/9$).

\section{Analysis Methodology and Results}
\label{analysis_sec_tom}
With the FESRs now defined in Eqs.~\eqref{eq:fesr-k-0}--\eqref{eq:fesr-k-2}, a lower bound on $a_\mu^\mathrm{QCD}$ can be constructed via \eqref{eq:a_mu_QCD-fesr_xi} [see also \eqref{eq:a_mu_QCD-fesr_summary}].  The methodology seeks to optimize $s_0$ such that it simultaneously maximizes the ratio $F_0^3/F_1^2$ to obtain the strongest possible bound, while still satisfying the inequality \eqref{eq:Cauchy-Schwarz} with $k=1$.  This ensures that the  resulting $s_0^\mathrm{opt}$ is in the region of validity for the FESRs because they satisfy the same inequality properties as an integrated hadronic spectral function. We start scanning $s_0$ from large energy (beginning near bottom threshold in $N_f=4$ regime) and find that stronger bounds trend toward lower $s_0$. We then transition to $N_f=3$ below the charm threshold (uncertainties associated with the Ref.~\cite{PDG2022}  value for the $m_c\left(m_c\right)=1.27\,{\rm GeV}$ threshold are negligible) .  

We use two different implementations of this  optimization methodology.  The flavour-separated approach applies the methodology to the FESRs (with each charge factor included) for each flavour separately, and then combines the individual optimized flavour contributions to obtain the final bound on $a_\mu^\mathrm{QCD}$.  In the flavour-combined approach, the methodology is applied to a combined FESR with a sum over flavours (with their  charge factors included).  The strongest bound from these two implementations is then used for our final prediction of the lower bound on $a_\mu^\mathrm{QCD}$.

We find that the flavour-separated approach leads to the strongest bound, and Table~\ref{result_tab_tom} shows the results for the central values of the QCD input parameters of Table~\ref{table:parameters}. There are a few key points in the interpretation of Table~\ref{result_tab_tom}.  First, it is important to remember that \eqref{eq:a_mu_QCD-fesr_xi} is truly a bound, and the optimized $s_0^\mathrm{opt}$ represents the value which maximizes the bound while simultaneously satisfying the $k=1$ inequality \eqref{eq:Cauchy-Schwarz}. It is therefore incorrect to interpret $s_0^\mathrm{opt}$ as a cut-off on the QCD contributions.  Second, the only field-theoretical distinction between the $u$ and $d$ contributions arises from the very small effect of quark masses, and hence  $s_0^\mathrm{opt}$ is the same in the non-strange channels and the bounds on $a_\mu^\mathrm{QCD}$ are in the ratio of quark charges $Q_u^2/Q_d^2=4$.  Third, the strange contributions to the $a_\mu^\mathrm{QCD}$ bound are roughly an order of magnitude smaller than non-strange, a feature that aligns with the data-driven and LQCD approaches to $a_\mu^{\mathrm{HVP,LO}}$ \cite{AOYAMA20201,PhysRevD.101.014029}. Finally,  we note that the entire inequality analysis of Section~\ref{sec:fesr} would also apply to Laplace sum-rules, leading to  analogous expressions for Eq.~\eqref{eq:a_mu_QCD-fesr_xi}.  We have explored this possibility and find that the Laplace sum-rule bounds are considerably weaker than for FESRs, presumably because the Laplace sum-rule  kernel $\exp(-t\tau)$ suppresses  higher-energy contributions compared to the polynomial FESR kernels.

\begin{table}[htb]
\centering
\renewcommand{\arraystretch}{1.5}
\begin{tabular}{|c|c|c|c|}
\hline
Flavour & $s_0^\mathrm{opt}\,\left(\mathrm{GeV}^2 \right)$ & $a_\mu^\mathrm{QCD}$ (lower bound)
& $a_\mu^\mathrm{QCD}$  (upper bound)\\ \hline
$u$ & $1.09$ & $\ge 472.7\times 10^{-10}$  & $\le 567.2\times 10^{-10} $
\\
\hline
$d$ & $1.09$ & $\ge 118.1\times 10^{-10}$  & $\le 141.7\times 10^{-10} $
\\
\hline
$s$ & $1.19$ & $\ge 66.2\times 10^{-10}$ & $\le 79.5\times 10^{-10} $
\\
\hline
Total & -- & $\ge 657.0\times 10^{-10}$ & $\le 788.4\times 10^{-10} $
\\\hline
\end{tabular}
\caption{The optimized $s^\mathrm{opt}_0$ and corresponding bounds on $a_\mu^\mathrm{QCD}$ are shown for each flavour in the flavour-separated method for central values of the QCD input parameters of Table~\ref{table:parameters}.  The total entry represents the sum of the individual flavour contributions for the final predicted bounds on $a_\mu^\mathrm{QCD}$.}
\label{result_tab_tom}
\end{table}

An uncertainty analysis was performed to determine the sensitivity of the Table~\ref{result_tab_tom} lower  $a_\mu^\mathrm{QCD}$ bounds arising from the QCD input parameters in Table~\ref{table:parameters}. The uncertainty of the   $a_\mu^\mathrm{QCD}$ bound is dominated by
changes in the vacuum saturation parameter  $\kappa$ and in the uncertainty of the dimension-four gluon
condensate parameter $\langle \alpha G^2\rangle$ (the poorly known strange-quark condensate parameter $r_c$ is 
a sub-dominant effect because the strange contributions in Table~\ref{result_tab_tom} are much smaller than  non-strange).  Taking into account the combined effect of these uncertainties gives our final QCD prediction for the light-quark contributions lower bound
\begin{equation}
a_\mu^\mathrm{QCD} \geq \left(657.0\pm 34.8\right)\times10^{-10}\,.
   \label{eq:a_mu_QCD-result_tom}
\end{equation}

A similar methodology is used to analyze the upper bounds associated with Eq.~\eqref{eq:a_mu_QCD-fesr_upper} [see also \eqref{eq:a_mu_QCD-fesr_summary}] using either \eqref{fm2_B_1} or  \eqref{fm2_B_2} for the upper bound on $F_{-2}$.  We seek the strongest bound that simultaneously satisfies the $k=1$ inequality \eqref{eq:Cauchy-Schwarz} along with the conditions \eqref{fm1_B_condition}, \eqref{fm2_B_condition_1}, and \eqref{fm2_B_condition_2}. As in the lower bound analysis, the flavour-separated approach leads to the strongest bound, and the same $s_0^{\rm opt}$ is obtained because the  $k=1$ Cauchy-Schwarz inequality \eqref{eq:Cauchy-Schwarz} turns out to be a limiting constraint in both cases. The results shown in Table~\ref{result_tab_tom} along with the theoretical uncertainty gives our final 
 QCD prediction for the light-quark contributions upper bound
\begin{equation}
a_\mu^\mathrm{QCD} \leq \left(788.4\pm 41.8\right)\times10^{-10}\,.
   \label{eq:a_mu_QCD-result_tom_upper}
\end{equation}

\begin{figure}[htb]
    \centering
    \includegraphics[scale=1]{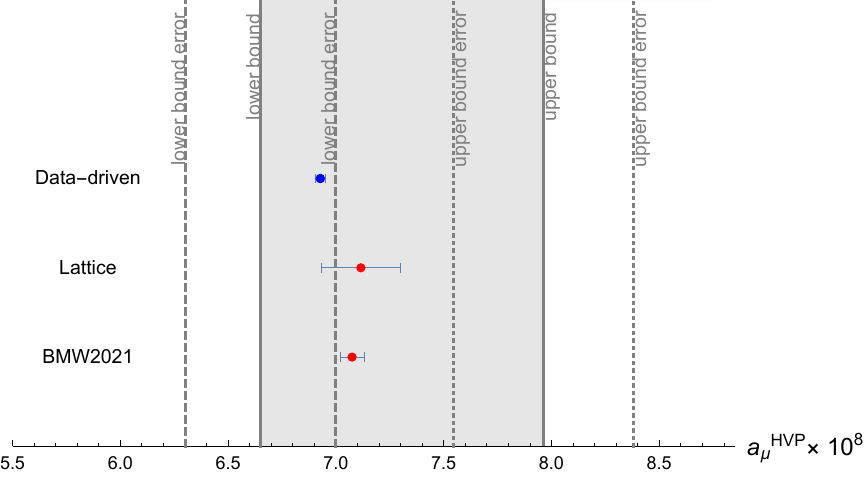}
    \caption{The $a_\mu^\mathrm{QCD}$ results \eqref{inclusive_bound_tom} showing lower bound (long dashed lines reflecting theoretical uncertainties) and upper bound (short dashed lines reflecting theoretical uncertainties) in comparison to 
    the $a_\mu^{\mathrm{HVP,LO}}$ world theoretical averages given in \cite{AOYAMA20201}. The blue indicates a data-driven methodology, while red indicates a value obtained via LQCD. Both the LQCD world average \cite{AOYAMA20201} and the sub-percent precision calculation from the BMW collaboration \cite{Borsanyi:2020mff} are shown for comparison.
    The grey shaded region illustrates the allowed central-value range of our QCD predictions in Eq.~\eqref{inclusive_bound_tom}.
    }
    \label{fig:comparison-jason-Apr02a}
\end{figure}

For purposes of comparison with data-driven approaches, we first note that although we are calculating light-quark contributions (and ultimately  using $N_f=3$ virtual corrections in the final results), our determinations \eqref{eq:a_mu_QCD-result_tom} and 
\eqref{eq:a_mu_QCD-result_tom_upper} still incorporate high-energy perturbative contributions to $a_\mu^\mathrm{QCD}$. We are thus underestimating the perturbative contributions above the charm threshold, and so our bounds remain valid.  Thus we have to supplement our bounds with charmonium and bottomonium resonance contributions of 
$a_{\mu\,,\,\bar{c}c\,,\,\bar{b}b}^{\mathrm{HVP,LO}}=(7.93\pm 0.19)\times 10^{-10}$ from \cite{PhysRevD.101.014029} to obtain  our total bound for comparison purposes 
\begin{equation}
 \left(664.9\pm 34.8\right)\times10^{-10}  \leq a_\mu^\mathrm{HVP,LO}  \leq 
 \left(796.3\pm 41.8\right)\times10^{-10}
 \label{inclusive_bound_tom}
\end{equation}
which should be compared with the data-driven Ref.~\cite{PhysRevD.101.014029} result
    \begin{equation}
    a_\mu^{\mathrm{HVP,LO}} = \left(692.78\pm 2.42\right)\times10^{-10},
    \label{eq:data-driven-result_tom}
\end{equation}
the data-driven result reported in the ($g-2$) Theory Initiative Whitepaper  \cite{AOYAMA20201} 
\begin{equation}
    a_\mu^{\mathrm{HVP,LO}} = \left(693.1\pm 4.0\right)\times10^{-10}, 
     \label{eq:data-driven-result_tom_2}
\end{equation}
as well as the result from LQCD reported in the ($g-2$) Theory Initiative Whitepaper \cite{AOYAMA20201},
\begin{equation}
    a_\mu^{\mathrm{HVP,LO}} = \left(711.6\pm 18.4\right)\times10^{-10}.
    \label{eq:lqcd-result_jason}
\end{equation}
These values can been seen compared against our bounds in Figure~\ref{fig:comparison-jason-Apr02a}.

In conclusion, we have constructed bounds on the 
QCD contributions to $a_\mu^{\mathrm{HVP,LO}}$ using 
a family of H\"older inequalities and related inequality constraints for QCD finite-energy sum-rules (FESRs).  These fundamental inequalities are based on the requirement that the QCD FESRs are consistent with the relation \eqref{eq:fesr} to an integrated hadronic spectral function, providing a novel methodology complementary to lattice QCD and data-driven approaches to determining $a_\mu^{\mathrm{HVP,LO}}$. Analyzing the light-quark ($u,d,s$) contributions up to five-loop order in perturbation theory in the chiral limit, LO in light-quark mass corrections, NLO in dimension-four QCD condensates,  and to LO in dimension-six QCD condensates leads to our QCD bounds in Eqs.~\eqref{eq:a_mu_QCD-result_tom} and \eqref{eq:a_mu_QCD-result_tom_upper}, which can be supplemented with the well-known contributions from charmonium and bottomonium states to obtain  the QCD bounds given in Eq.~\eqref{inclusive_bound_tom}.  As shown in the Appendix, these FESR bounds are more restrictive than the updated Laplace sum-rule bounds using the approach of  Ref.~\cite{Steele1991}. 
As illustrated in Fig.~\ref{fig:comparison-jason-Apr02a}, the central values of our total QCD bounds \eqref{inclusive_bound_tom} thus bridge the region between LQCD and data-driven values, indicating 
a possible resolution of the tension between LQCD and data-driven determinations of $a_\mu^{\mathrm{HVP,LO}}$. Resolving this tension would provide better guidance to searches for new physics in measurements of the anomalous magnetic moment of the muon. In future work we will search for new methods and new fundamental inequalities to improve bounds on the QCD contributions to $a_\mu^{\mathrm{HVP,LO}}$. 

\section*{Acknowledgments}
TGS is grateful for research funding from the Natural Sciences and Engineering Research Council of Canada (NSERC).

\section*{Appendix: Laplace Sum Rule Approach}\label{sec:lsr}
QCD Laplace sum-rules \cite{Shifman:1978bx,Shifman:1978by} are similar to finite-energy sum-rules as defined in~\eqref{eq:fesr}; however, they are constructed using a Borel (inverse Laplace) transform which introduces an exponential factor:
\begin{equation}
L_k\left(\tau\,,s_0\right) = \int_{t_0}^{s_0}\frac{1}{\pi}\mathrm{Im}\Pi^H \left(t\right)  t^k \, e^{-t \tau}\, \mathrm{d}t\,.
\label{eq:lsr}
\end{equation}
In \cite{Steele1991} it was shown that $a_\mu^{\mathrm{HVP,LO}}$, as defined in \eqref{eq:a_QCD}, can be expressed as a linear combination of QCD Laplace sum-rules \eqref{eq:lsr}. First, the exact kernel function~\eqref{eq:K_exact} can be approximated near $t=t^\prime$ as
\begin{equation}
K\left(t\right) \approx \mathcal{K}\left(t\,,t^\prime\right) = K\left(t^\prime\right) e^\zeta \left[a_1 \left(\frac{t}{t^\prime}\right) + a_2 \left(\frac{t}{t^\prime}\right)^2 + a_3 \left(\frac{t}{t^\prime}\right)^3\right] e^{-\zeta t/t^\prime}\,,
    \label{eq:K_lsr_approx}
\end{equation}
where $a_1+a_2+a_3=1$ so that $K\left(t^\prime\right) = \mathcal{K}\left(t^\prime\,,t^\prime\right)$. 
Inserting \eqref{eq:K_lsr_approx} into \eqref{eq:a_QCD} yields 
\begin{equation}
    a_\mu^\mathrm{QCD} \approx 4\alpha^2 \, K\left(t^\prime\right)\frac{e^\zeta}{t^\prime}\int_{t_0}^\infty  
    \frac{1}{\pi} \mathrm{Im}\Pi^H\left(t\right) \left[a_1 + a_2 \left(\frac{t}{t^\prime}\right) + a_3 \left(\frac{t}{t^\prime}\right)^2\right] e^{-\zeta t/t^\prime} \mathrm{d}t\,,
    \label{eq:a_mu_lsr_0}
\end{equation}
where $t_0=4m_\pi^2$. Introducing the parameter $s_0$ as in \eqref{eq:a_mu_QCD_tom} and defining $\tau = \zeta/t^\prime$, \eqref{eq:a_mu_lsr_0} becomes
\begin{equation}
    a_\mu^\mathrm{QCD} \approx 4\alpha^2 \, K\left(\zeta/\tau\right)\frac{\tau}{\zeta} e^\zeta \int_{t_0}^{s_0}  
    \frac{1}{\pi} \mathrm{Im}\Pi^H\left(t\right) \left[a_1 + a_2 \left(\frac{t}{t^\prime}\right) + a_3 \left(\frac{t}{t^\prime}\right)^2\right] e^{-t\tau} \mathrm{d}t\,.
    \label{eq:a_mu_lsr_1}
\end{equation}
Comparing \eqref{eq:a_mu_lsr_1} and the definition of the Laplace sum-rules in \eqref{eq:lsr} shows that we may approximate $a_\mu^{\mathrm{HVP,LO}}$ as a linear combination of Laplace sum-rules:
\begin{equation}
a_\mu^\mathrm{QCD} \approx 4\alpha^2 \, K\left(\zeta/\tau\right)\frac{\tau}{\zeta} e^\zeta \left[a_1 L_0\left(\tau\,,s_0\right)+a_2\frac{\tau}{\zeta}L_1\left(\tau\,,s_0\right)+a_3\left(\frac{\tau}{\zeta}\right)^2 L_2\left(\tau\,,s_0\right)\right]\,.
\label{eq:a_mu_lsr_2}
\end{equation}
The approximation \eqref{eq:K_lsr_approx} is used because it makes a theoretical calculation of $a_\mu^{\mathrm{HVP,LO}}$ (using a QCD expression for the vacuum polarization function) amenable to a Laplace sum-rule analysis. In \eqref{eq:K_lsr_approx} the expansion is truncated at $\mathcal{O}(t^3)$ to avoid dependence on unknown higher dimension QCD condensates (a similar issue is encountered in the finite-energy sum-rule analysis in Section~\ref{sec:fesr}). 

\begin{figure}[htb]
\centering
    \includegraphics[width=0.5\textwidth]{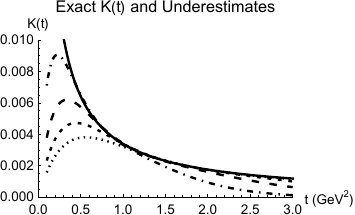}\includegraphics[width=0.5\textwidth]{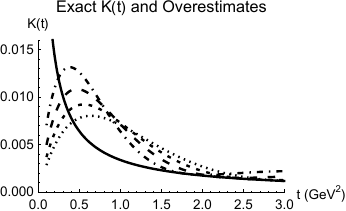}
    \caption{Left: the exact $K(t)$ (solid line) compared to underestimates $\mathcal{K}^\downarrow\left(t\,,t^\prime\right)$ with $t^\prime\in\left\{0.8\,,\,1.2\,,\,1.6\,,\,2.0\right\}\,{\rm GeV^2}$, which are respectively represented by the dashed dotted, long dashed, short dashed, and dotted lines. Right: the exact $K(t)$ (solid line) compared to overestimates $\mathcal{K}^\uparrow\left(t\,,t^\prime\right)$ with $t^\prime\in\left\{1.8\,,\,2.2\,,\,2.6\,,\,3.0\right\}\,{\rm GeV^2}$, which are respectively represented by dashed dotted, long dashed, short dashed, and dotted lines. The parameters  used in equation~\eqref{eq:K_lsr_approx} for the underestimates  $\mathcal{K}^\downarrow\left(t\,,t^\prime\right)$ ($\{a_1=1.5700,\, a_2=-1.75658,\, a_3=1.1958,\, \zeta=2.6528\}$) and overestimates  $\mathcal{K}^\uparrow\left(t\,,t^\prime\right)$ ($\{a_1=6.0378,\, a_2=-10.7006,\, a_3=5.6628,\, \zeta=2.6528\}$) are identical to those used in Ref.~\cite{Steele1991}.}
    \label{K_under_over_fig}
\end{figure}

Although the approximation \eqref{eq:K_lsr_approx} is designed to be exact at $t=t^\prime$  and is well suited to a Laplace sum-rule analysis, the approximation of the exact kernel function \eqref{eq:K_exact} decreases in accuracy away from $t=t^\prime$. In order to gain some control over the theoretical uncertainty introduced by this approximation we will follow the approach of Ref.~\cite{Steele1991}, wherein the approximation \eqref{eq:K_lsr_approx} was used to construct underestimates and overestimates of the exact kernel function \eqref{eq:K_exact}, respectively denoted as $\mathcal{K}^\downarrow\left(t\,,t^\prime\right)$  (corresponding to parameters $\{a_1=1.5700,\, a_2=-1.75658,\, a_3=1.1958,\, \zeta=2.6528\}$) and $\mathcal{K}^\uparrow\left(t\,,t^\prime\right)$ (corresponding to parameters $\{a_1=6.0378,\, a_2=-10.7006,\, a_3=5.6628,\, \zeta=2.6528\}$), which are shown in Fig.~\ref{K_under_over_fig}. Using these underestimates and overestimates, a QCD Laplace sum-rule analysis can be performed to generate lower and  upper bounds on $a_\mu^{\mathrm{HVP,LO}}$.

Using the results of Eqs.~\eqref{eq:correlator} and \eqref{eq:pertseries}, the Laplace sum-rules (LSRs)
for light-quark ($u,d,s$) contributions up to five-loop order in perturbation theory in the chiral limit, LO in light-quark mass corrections, next-to-leading order (NLO) in dimension-four QCD condensates,  and to LO in dimension-six QCD condensates are given for a generic light flavour by
{\allowdisplaybreaks
\begin{gather}
\begin{split}
    L_0\left(\tau,s_0\right)=&\frac{1}{4\pi^2\tau}\left[ 
     f_{0,0}\left(\tau s_0\right)  +
    \sum_{k=0}^3 f_{0,k}\left(\tau s_0\right) \sum_{j=k+1}^4 T_{j,k} \left( \frac{\alpha_s(\nu)}{\pi}\right)^j
    \right]-\frac{3}{2\pi^2}m_q(\nu)^2
    \\
   & + 2\langle m_q \bar{q}q\rangle\left(1+\frac{1}{3}\frac{\alpha_s(\nu)}{\pi}\right)\tau
   + \frac{1}{12\pi}\langle \alpha_s G^2 \rangle\left(1+\frac{7}{6}\frac{\alpha_s(\nu)}{\pi}\right)\tau
   - \frac{112}{81} \pi \alpha_s \langle \bar{q}\bar{q}qq\rangle\tau^2\,,
    \end{split}
    \label{L0}
\\
 \begin{split}
    L_1\left(\tau,s_0\right)=&\frac{1}{4\pi^2\tau^2}\left[ 
     f_{1,0}\left(\tau s_0\right)  +
    \sum_{k=0}^3 f_{1,k}\left(\tau s_0\right) \sum_{j=k+1}^4 T_{j,k} \left( \frac{\alpha_s(\nu)}{\pi}\right)^j
    \right]
   \\
   & - 2\langle m_q \bar{q}q\rangle\left(1+\frac{1}{3}\frac{\alpha_s(\nu)}{\pi}\right)
   - \frac{1}{12\pi}\langle \alpha_s G^2 \rangle\left(1+\frac{7}{6}\frac{\alpha_s(\nu)}{\pi}\right)
   + \frac{224}{81} \pi \alpha_s \langle \bar{q}\bar{q}qq\rangle\tau\,,
    \end{split}  
    \label{L1}
    \\
    L_2\left(\tau,s_0\right)=\frac{1}{4\pi^2\tau^3}\left[ 
     f_{2,0}\left(\tau s_0\right)  +
    \sum_{k=0}^3 f_{2,k}\left(\tau s_0\right) \sum_{j=k+1}^4 T_{j,k} \left( \frac{\alpha_s(\nu)}{\pi}\right)^j
    \right]
   - \frac{224}{81} \pi \alpha_s \langle \bar{q}\bar{q}qq\rangle\,,
\label{L2}  
\end{gather}
}
where we have defined the quantity
\begin{equation}
f_{j,k}\left(\tau s_0\right)=\int_0^{\tau s_0} z^j\left[\log{\left(\frac{1}{z}\right)} \right]^k e^{-z} \,dz\,.
\label{f_jk_eq}
\end{equation} 
Implicit in Eqs.~\eqref{L0}--\eqref{L2} is a renormalization scale of $\nu = 1/\sqrt{\tau}$ in both $\alpha_s$ and the running quark masses \cite{Narison:1981ts}. 
As in the QCD expressions  \eqref{eq:fesr-k-0}--\eqref{eq:fesr-k-2} for the FESRs, the generic light-flavour LSRs  \eqref{L0}--\eqref{L2} require a pre-factor of their quark charge.

\begin{figure}[htb]
\centering
    \includegraphics{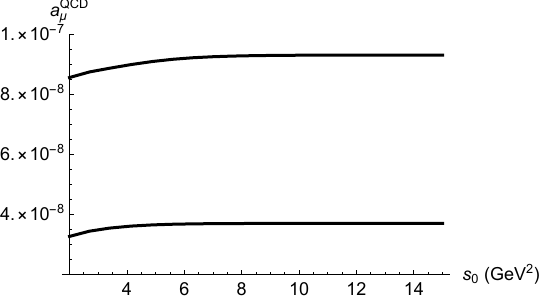}
    \caption{LSR upper bound (top curve) and lower bound (bottom curve) 
    on light-quark contributions to $a_\mu^{\rm QCD}$ as a function of $s_0$.
    }
    \label{LSR_fig}
\end{figure}

Following the analysis methodology Ref.~\cite{Steele1991} for determining the upper and lower bounds on $a_\mu^{\rm QCD}$,  $\tau$ stability \cite{Bell:1980ub,Bell:1980ww,Narison:2023ntg} is used to determine the right-hand side of \eqref{eq:a_mu_lsr_2} for a fixed $s_0$, and then $s_0$ is varied until an asymptotic value is reached. 
The $\tau$-stability region naturally tends toward the $N_f=3$ regime.
As with the FESRs, this methodology can be applied to either a flavour-separated or flavour-combined case, but unlike the FESRs there is negligible difference in the two cases. Fig.~\ref{LSR_fig} shows the results for  central values of the QCD input parameters, and leads to the bounds
\begin{equation}
 369.5\times 10^{-10}\le a_\mu^{\rm QCD}\le 930.2\times 10^{-10}\,.
    \label{LSR_bounds}
\end{equation}
 Comparing Eq.~\eqref{LSR_bounds} with the FESR results in Eqs.~\eqref{eq:a_mu_QCD-result_tom} and \eqref{eq:a_mu_QCD-result_tom_upper} it is evident that the FESR bounds are stronger than those obtained from updated and extended  QCD inputs in the  Ref.~\cite{Steele1991} LSR methodology.

\bibliographystyle{h-physrev}
\bibliography{g-2.bib}
\end{document}